\documentclass[aps,twocolumn,showpacs,floatfix]{revtex4}
\usepackage{amssymb,amsbsy,psfig,graphicx,bbm}
\newcommand{\gr}[1]{\boldsymbol{#1}}
\newcommand{\be}{\begin{equation}}
\newcommand{\ee}{\end{equation}}
\newcommand{\bea}{\begin{eqnarray}}
\newcommand{\eea}{\end{eqnarray}}

\newcommand{\N}{{\cal N}}

\begin{document}
\title{Entanglement and purity of two--mode Gaussian states in noisy channels}
\author{Alessio Serafini$^{1}$}\email{serale@sa.infn.it}
\author{Fabrizio Illuminati$^{1}$}\email{illuminati@sa.infn.it}
\author{Matteo G. A. Paris$^{2,3}$}\email{matteo.paris@fisica.unimi.it}
\author{Silvio De Siena$^{1}$}\email{desiena@sa.infn.it}
\affiliation{
\mbox{\mbox{}$^{1}$Dipartimento di Fisica ``E. R. Caianiello'',
Universit\`a di Salerno, INFM UdR Salerno,}\\
\mbox{INFN Sezione Napoli, Gruppo Collegato Salerno,
Via S. Allende, 84081 Baronissi (SA), Italia} \\
\mbox{}$^{2}$ Dipartimento di Fisica, Universit\'a di Milano, Italia.
\\ \mbox{}$^{3}$ Dipartimento di Fisica ``A. Volta'', Universit\`a di Pavia,
Italia.} 
\begin{abstract}
We study the evolution of purity, entanglement and total correlations 
of general two--mode continuous variable Gaussian states
in arbitrary uncorrelated Gaussian environments. The time evolution 
of purity, Von Neumann entropy, logarithmic negativity and 
mutual information is analyzed for a wide range of initial conditions. 
In general, we find that a local squeezing of the bath
leads to a faster degradation of purity and entanglement, while
it can help to preserve the mutual information between the modes. 
\end{abstract}
\date{Februar 06, 2004} 
\pacs{3.67.-a, 3.67.Pp, 42.50.Dv}
\maketitle
\section{Introduction}
In recent years, it has been increasingly realized that
Gaussian states and Gaussian channels are
essential ingredients of continuous variable quantum 
information \cite{pati03}. Indeed, entangled Gaussian states have been 
successfully exploited in realizations of quantum key 
distribution \cite{crypto} and teleportation \cite{tele} protocols.\par
In such experimental settings, 
the entanglement of a bipartite state is usually 
distilled locally, and then distributed over space, letting  
the entangled subsystems evolve independently 
and move to separated spatial regions. In the course
of this processes, interaction with the external
environment is unavoidable and must be properly
understood.
Therefore, the analysis of the evolution of quantum  
correlations and decoherence of Gaussian states in
noisy channels is of crucial interest, and has 
spurred several theoretical 
works \cite{duan97,hiroshima01,scheel01,paris02,kim03,paris04,prauz03}.\par
The evolution of fidelity of generic bosonic fields  
in noisy channels has been addressed in Ref.~\cite{duan97}. 
Indeed, the relevant instance of initial 
two--mode squeezed vacua (possessing nontrivial entanglement 
properties) has drawn most of the attention in the field.
Decoherence and entanglement degradation of such states 
in thermal baths have been analyzed in
Refs.~\cite{scheel01,paris02}, whereas phase damping
and the effects of squeezed reservoirs are dealt with 
in Refs.~\cite{hiroshima01,kim03,paris04}. 
In Ref.~\cite{prauz03} the author studies the evolution
of a two--mode squeezed vacuum 
in a common bath endowed with cross correlations and 
asymptotic entanglement. 
Decoherence and entanglement degradation in 
continuous variable systems have been experimentally 
investigated in Ref.~\cite{bowen03}.\par
In this paper we address the general case of an arbitrary
two--mode Gaussian state dissipating in arbitrary local
Gaussian environments. 
The resulting dynamics is governed by 
a two--mode master equation describing losses and 
thermal hopping in presence of (local) non classical 
fluctuations of the environment.\par
We study the evolution of quantum and total correlations 
and the behavior of decoherence in noisy channels.  
Quantum and total correlations of a state 
will be quantified by, respectively, 
its logarithmic negativity \cite{vidwer} and its mutual information, 
while the rates of decoherence will be determined by following  
the evolution of the purity (conjugate to the linear entropy) and of the
Von Neumann entropy. We present explicit analytic results, as well as 
numerical studies, on the optimization of the relevant physical quantities 
along the non-unitary evolution. Our analysis provides an answer to 
the question whether possible effective schemes to mimic general 
Gaussian environments \cite{tomvit,cir} 
are able to delay the decay of quantum 
coherence and correlations. 
We mention that, among such schemes, the most interesting 
for applications to bosonic fields is based 
on quantum non demolition (QND)
measurements and feedback dynamics \cite{tomvit,wise}. 
We finally remark that the optimization of the quantities we are going 
to study with respect to phenomenological parameters turns out to be 
particularly relevant at `small times', before decoherence has 
irreversibly corrupted the quantum features of the state, crucial for 
applications in quantum information.\par 
This paper is structured as follows.
In Section II we provide a self-contained description of the general
structure of two-mode Gaussian states, including the characterization
of their mixedness and entanglement. In Section III we review 
the evolution of Gaussian states in general
Gaussian environments. In Section IV we focus on the evolution of
purity and entanglement, determining the optimal regimes that
can help preserving these quantities from environmental corruption.
Finally, in Section V we summarize our results and discuss some
outlook on future research.   
\section{Two--mode Gaussian states: general properties\label{sec2m}}
Let us consider a two--mode continuous variable system, described by an 
Hilbert space ${\cal H}={\cal H}_{1}\otimes{\cal H}_{2}$ resulting from the tensor product 
of the Fock spaces ${\cal H}_{i}$'s. 
We denote by $a_{i}$ the annihilation operator acting on the space ${\cal H}_{i}$,
and by $\hat x_{i}=(a_{i}+a^{\dag}_{i})/\sqrt{2}$ and
$\hat p_{i}=(a_{i}-a^{\dag}_{i})/\sqrt{2}$
the quadrature phase operators related to the mode $i$ of the field.
The corresponding phase space variables will be denoted by $x_{i}$ and $p_{i}$.\par
The set of Gaussian states is, by definition, the set of states with Gaussian characteristic 
functions and quasi--probability distributions. Therefore a Gaussian state is completely
characterized by its first and second statistical moments, which will be denoted, respectively, 
by the vector of first moments $\bar X\equiv\left(\langle\hat x_{1}
\rangle,\langle\hat p_{1}\rangle,\langle\hat x_{2}\rangle,
\langle\hat p_{2}\rangle\right)$ 
and by the covariance matrix $\boldsymbol{\sigma}$
\begin{equation}
\sigma_{ij}\equiv\frac{1}{2}\langle \hat{x}_i \hat{x}_j + 
\hat{x}_j \hat{x}_i \rangle -
\langle \hat{x}_i \rangle \langle \hat{x}_j \rangle \, .
\end{equation}
First moments can be arbitrarily adjusted by local unitary operations, which 
do not affect any quantity related to entanglement or mixedness.
Moreover, as we will see in Sec.~\ref{evo}, 
they do not influence the evolution of second moments 
in the instances we will deal with. 
Therefore they will be unimportant to our aims and we will set them
to $0$ in the following, without any loss of generality for our subsequent results.
Throughout the paper, $\gr{\sigma}$ will stand both for the covariance matrix 
and the Gaussian state $\varrho$ itself.\par
It is convenient to express $\boldsymbol{\sigma}$ in terms of the three $2\times 2$
matrices $\boldsymbol{\alpha}$, $\boldsymbol{\beta}$, $\boldsymbol{\gamma}$
\begin{equation}
\boldsymbol{\sigma}\equiv\left(\begin{array}{cc}
\boldsymbol{\alpha}&\boldsymbol{\gamma}\\
\boldsymbol{\gamma}^{T}&\boldsymbol{\beta}
\end{array}\right)\, . \label{espre}
\end{equation}
Positivity of $\varrho$ and the canonical commutation relations
impose the following constraint for $\boldsymbol{\sigma}$ to
be a {\em bona fide} covariance matrix \cite{simon00}
\begin{equation}
\boldsymbol{\sigma}+\frac{i}{2}\boldsymbol{\Omega}\ge 0 \; ,
\label{bonfide}
\end{equation}
where $\boldsymbol{\Omega}$ is the standard symplectic form
\[
\boldsymbol{\Omega}\equiv \left(\begin{array}{cc}
\boldsymbol{\omega}&0\\
0&\boldsymbol{\omega}
\end{array}\right)\; , \quad \boldsymbol{\omega}\equiv \left( \begin{array}{cc}
0&1\\
-1&0
\end{array}\right) \; .
\]
Inequality (\ref{bonfide}) is a useful and elegant 
way to express Heisenberg uncertainty principle.\par
In the following, we will make use of the Wigner 
quasi--probability representation
$W$, defined as the Fourier transform of the 
symmetrically ordered characteristic function \cite{barnett}. 
In Wigner phase space picture, the tensor product 
${\cal H}={\cal H}_{1}\otimes{\cal H}_{2}$ of the Hilbert 
spaces $H_{i}$'s of the two modes results in the direct sum 
$\Gamma=\Gamma_{1}\oplus\Gamma_{2}$ of the associated phase spaces 
$\Gamma_{i}$'s. 
A symplectic transformation acting on the 
global phase space $\Gamma$ corresponds to a unitary operator acting on the
global Hilbert space $\cal H$ \cite{simon8794}. 
In what follows we will refer to a transformation $S_{l} = S_{1} \oplus S_{2}$, with each 
$S_{i} \in Sp_{(2,\mathbb R)}$ acting on 
$\Gamma_{i}$, as to a ``local symplectic operation''.
The corresponding unitary transformation is 
the ``local unitary transformation'' $U_{l}=
U_{1}\otimes U_{2}$, with each $U_{i}$ acting on ${\cal H}_{i}$.\par
The Wigner function of a Gaussian state 
can be written as follows in terms of phase space quadrature variables
\begin{equation}
W(X)=\frac{\,{\rm e}^{-\frac{1}{2}X\boldsymbol{\sigma}^{-1}X^{T}}}{\pi\sqrt{{\rm
Det} [\boldsymbol{\sigma}]}}{\:,}\label{wigner}
\end{equation}
where $X$ stands for the vector $(x_{1},p_{1},x_{2},p_{2})\in\Gamma$. \par
It is well known that for any covariance matrix $\boldsymbol{\sigma}$ there exists a local 
canonical operation $S_{l}=S_{1}\oplus S_{2}$ which transforms 
$\boldsymbol{\sigma}$ to the so called standard form $\boldsymbol{\sigma}_{sf}$ \cite{duan00}
\begin{equation}
S_{l}^{T}\boldsymbol{\sigma}S_{l}=\boldsymbol{\sigma}_{sf}
\equiv \left(\begin{array}{cccc}
a&0&c_{1}&0\\
0&a&0&c_{2}\\
c_{1}&0&b&0\\
0&c_{2}&0&b
\end{array}\right)\; . \label{stform}
\end{equation}
States whose standard form fulfills $a=b$ are said to be symmetric.
Let us recall that any pure state is symmetric and fulfills 
$c_{1}=-c_{2}=\sqrt{a^2-1/4}$. 
The correlations $a$, $b$, $c_{1}$, and $c_{2}$ are determined by the four local symplectic 
invariants ${\rm Det}\boldsymbol{\sigma}=(ab-c_{1}^2)(ab-c_{2}^2)$, 
${\rm Det}\boldsymbol{\alpha}=a^2$, ${\rm Det}\boldsymbol{\beta}=b^2$, 
${\rm Det}\boldsymbol{\gamma}=c_{1}c_{2}$. Therefore, the standard form 
corresponding to any covariance matrix is unique.  \par
Inequality (\ref{bonfide}) can be recast as a constraint on 
the $Sp_{(4,{\mathbb R})}$ invariants ${\rm Det}\gr{\sigma}$ and 
$\Delta(\gr{\sigma})={\rm Det}\boldsymbol{\alpha}+\,{\rm Det}\boldsymbol{\beta}+2
\,{\rm Det}\boldsymbol{\gamma}$:
\begin{equation}
\Delta(\gr{\sigma})\le\frac{1}{4}+4\,{\rm Det}\boldsymbol{\sigma}
\label{sepcomp}\; .
\end{equation}\par
Finally, let us recall that a centered two--mode Gaussian state can always be 
written as \cite{holevo,serafozzi}
\begin{equation}
\gr{\sigma}=S^T \gr{\nu} S \; ,
\end{equation}
where $S\in Sp_{(4,\mathbb{R})}$ and $\gr{\nu}$ is the tensor product of 
thermal states with covariance matrix
\begin{equation}
\gr{\nu}=\,{\rm diag}({n}_{-},{n}_{-},{n}_{+},{n}_{+}) \, .
\label{therma}
\end{equation}
The quantities $n_{\mp}$ form the symplectic spectrum of 
the covariance matrix $\gr{\sigma}$. They can be easily computed in terms of
the $Sp_{(4,\mathbb{R})}$ invariants
\begin{equation}
2{n}_{\mp}^2=\Delta(\gr{\sigma})\mp\sqrt{\Delta(\gr{\sigma})^2
-4\,{\rm Det}\,\gr{\sigma}} \, . \label{sympeig}
\end{equation}
The symplectic eigenvalues $n_{\mp}$ encode essential informations 
about the Gaussian state $\gr{\sigma}$ and 
provide powerful, simple ways to express its fundamental properties. 
For instance, the Heisenberg uncertainty relation (\ref{bonfide}) 
can be recast in the compact, equivalent form 
\be
n_{-}\ge\frac12 \: . \label{heisymp}
\ee
\par
A relevant subclass of Gaussian states we will make use of is 
constituted by the two--mode squeezed thermal states. Let 
$S_{r}=\exp(\frac12 r a_{1}a_{2}-\frac12 r a_{1}^{\dag}a_{2}^{\dag})$ 
be the two mode squeezing operator
with real squeezing parameter $r$, and let $\gr{\nu}_{\mu}=
1/(2\sqrt{\mu}){\mathbbm 1}$ be the 
tensor product of identical thermal states, where 
$\mu = {\rm Tr} \left( \varrho^{2}\right)$ is the purity of the state. 
Then, for a two-mode squeezed thermal state $\gr{\xi}_{\mu,r}$ 
we can write $\gr{\xi}_{\mu,r}=S_{r}\gr{\nu}_{\mu}S^{\dag}$. 
The covariance matrix of $\gr{\xi}_{\mu,r}$ is a symmetric standard form
satisfying
\be
a=\frac{\cosh2r}{2\sqrt{\mu}}\; , \quad 
c_{1}=-c_{2}=\frac{\sinh2r}{2\sqrt{\mu}} \, , \label{sqthe}
\ee
and in the instance $\mu=1$ one recovers the pure
two--mode squeezed vacuum states. 
Two--mode squeezed states are endowed with remarkable properties 
related to entanglement \cite{2max,giedke03}; their dynamics
in noisy channels will be analyzed in detail.
\subsection{Characterization of mixedness}
Let us briefly recall that the degree of purity of a quantum state can be properly characterized 
either by the Von Neumann entropy $S_{V}$ or by the linear entropy $S_{l}$. Such quantities are defined as 
follows for continuous variable systems
\begin{eqnarray}
S_{V}&\equiv&-\;{\rm Tr}(\varrho\;\ln\varrho)\; , \label{vneu}\\
S_{l}&\equiv&1-\;{\rm Tr}(\varrho^{2})\equiv1-\mu\; ,\label{linea}
\end{eqnarray}
where the purity $\mu\equiv\;{\rm Tr}(\varrho^{2})$ has already been
introduced . 
We first point out that $\mu$ can be easily computed for Gaussian states. In fact,
in the Wigner phase space picture the trace of a product of operators corresponds to the 
integral of the product of their Wigner representations (when existing) 
over the whole phase space. Because 
the representation of a state $\varrho$ is just $W$, 
for an $n$--mode Gaussian state we have, taking into account the proper 
normalization factor,
\begin{equation}
\mu(\gr{\sigma})=\frac{\pi}{2^{n}}\int_{{\mathbb R}^{2n}}
W^2\,{\rm d}^{n}x\,{\rm d}^{n}p=
\frac{1}{2^n \sqrt{\,{\rm Det}\,\boldsymbol{\sigma}}}\, .
\label{purezza}
\end{equation}\par
For Gaussian states, 
the Von Neumann entropy can be computed as well, 
determining their symplectic spectra. For single--mode 
Gaussian states, one has \cite{agarwal71}
\begin{equation}
S_{V}(\gr{\sigma}) = 
\frac{1-\mu}{2\mu}\ln\left(\frac{1+\mu}{1-\mu}\right)
-\ln\left(\frac{2\mu}{1+\mu}\right) \label{vneu1}\, ,
\end{equation}
where $\mu$ can be computed from Eq.~(\ref{purezza}) for $n=1$. 
$S_V$ is in this case 
an increasing function of the linear entropy, so that both 
quantities provide the same characterization
of mixedness. This is no longer true for two--modes Gaussian states: 
in this case the Von Neumann entropy reads \cite{holevo,serafozzi}
\begin{equation}
S_{V}(\gr{\sigma})=f[\tilde{n}_{-}
(\boldsymbol{\sigma})]+f[\tilde{n}_{+}(\boldsymbol{\sigma})]\; ,
\label{vneu2}
\end{equation}
where 
\[
f(x) \equiv (x+\frac12)\ln(x+\frac12)-(x-\frac12)\ln(x-\frac12)
\]
and the symplectic eigenvalues $n_{\mp}(\gr{\sigma})$ are 
given by Eq.~(\ref{sympeig}).\par
Knowledge of the Von Neumann entropy $S_{V}$ allows for the determination of the 
mutual information $I$ defined,  
for a general bipartite quantum state $\varrho$, 
as $I(\varrho)=S_{V}(\varrho_{1})+S_{V}(\varrho_{2})-S_{V}(\varrho)$, where 
$\varrho_{i}$ refers to the reduced state obtained tracing over the variables of
subsystem $j\neq i$. The mutual information $I(\gr{\sigma})$ of a 
two--mode Gaussian state $\gr{\sigma}$ reads \cite{serafozzi}
\begin{equation}
I(\gr{\sigma})=f(a)+f(b)-f(n_{-})-f(n_{+}) \; .
\label{muinf}
\end{equation}
One can make use of such a quantity to estimate the amount of 
total (quantum plus classical)
correlations contained in a state $\gr{\sigma}$ \cite{vedral01}.
\subsection{Characterization of entanglement}
We now review some properties of entanglement for two--mode Gaussian states.
The necessary and sufficient separability criterion for such states is  
positivity of the partially transposed state $\tilde{\gr{\sigma}}$ 
(``PPT criterion'') \cite{simon00}. 
It can be easily
seen from the definition of $W(X)$ that the action of partial transposition 
amounts, in phase space, to a mirror reflection of one of the four canonical variables. 
In terms of $Sp_{2,\mathbb{R}}\oplus Sp_{2,\mathbb{R}}$ invariants, 
this results in flipping the sign of ${\rm Det}\,\gr{\gamma}$. Therefore 
the invariant $\Delta(\gr{\sigma})$ is changed into $\tilde{\Delta}({\gr\sigma})
=\Delta(\tilde{\gr{\sigma}})=\,{\rm Det}\,\gr{\alpha}+
\,{\rm Det}\,\gr{\beta}-2\,{\rm Det}\,\gr{\gamma}$. Now, the symplectic
eigenvalues $\tilde{n}_{\mp}$ of $\tilde{\gr{\sigma}}$ read
\be
\tilde{n}_{\mp}=
\sqrt{\frac{\tilde{\Delta}(\gr{\sigma})\mp\sqrt{\tilde{\Delta}(\gr{\sigma})^2
-4\,{\rm Det}\,\gr{\sigma}}}{2}} \, . \label{sympareig}
\ee
The PPT criterion then reduces to a simple inequality that must
be satisfied by the smallest symplectic eigenvalue $\tilde{n}_{-}$
of the partially transposed state
\be
\tilde{n}_{-}\ge \frac12 \: ,
\label{symppt}
\ee
which is equivalent to 
\be
\tilde{\Delta}(\gr{\sigma})\le 4\,{\rm Det}\,\gr{\sigma}+\frac14 \; .
\label{ppt}
\ee 
The above inequalities
imply ${\rm Det}\,\gr{\gamma}=c_{1}c_{2}<0$ 
as a necessary condition for a two--mode Gaussian state 
to be entangled.
The quantity $\tilde{n}_{-}$ encodes all the qualitative characterization of 
the entanglement for arbitrary (pure or mixed) two--modes Gaussian states. 
Note that 
$\tilde{n}_{-}$ takes a particularly simple form for 
entangled symmetric states, whose standard 
form has $a=b$
\be
\tilde{n}_{-}=\sqrt{(a-|c_{1}|)(a-|c_{2}|)} \; .
\label{symeig}
\ee
\par
As for the quantification of entanglement, no 
fully satisfactory measure is known 
at present for arbitrary mixed two--mode Gaussian states.
However, a quantification of entanglement which can be 
computed for general two--mode Gaussian states is provided by the negativity
$\N$, introduced by Vidal and Werner for continuous variable systems \cite{vidwer}. 
The negativity of a quantum state 
$\varrho$ is defined as
\be
{\cal N}(\varrho)=\frac{\|\tilde \varrho \|_{1}-1}{2}\: ,
\ee 
where $\tilde\varrho$ is the partially transposed density matrix and 
$\|\hat o\|_{1}\equiv\,{\rm Tr}\,\sqrt{\hat o^{\dag}\hat o}$ 
stands for the trace norm of an operator $\hat o$. The quantity ${\cal N}
(\varrho)$ is equal to $|\sum_{i}\lambda_{i}|$, the modulus of the sum 
of the negative eigenvalues of $\tilde\varrho$, and it quantifies
the extent to which $\tilde\varrho$ fails to be positive. 
Strictly related to $\N$ is the logarithmic negativity $E_{\N}$, 
defined as $E_{\N}\equiv \ln\|\tilde{\varrho}\|_{1}$.
The negativity has been proved to be  
convex and monotone under LOCC (local operations and classical
communications) \cite{footnote1}, but fails to be 
continuous in trace norm on infinite dimensional 
Hilbert spaces. Anyway, this problem can 
be somehow eluded by restricting to states with finite 
mean energy \cite{jensplenio}.
For two--mode Gaussian states it can be easily shown that 
the negativity is a simple function of $\tilde{n}_{-}$, which is 
thus itself an (increasing) entanglement monotone; 
one has in fact \cite{vidwer}
\be
E_{\N}(\gr{\sigma})=\max\left\{0,-\ln{2\tilde{n}_{-}}\right\} \: .
\ee
This is a decreasing function of the smallest partially transposed symplectic 
eigenvalue $\tilde{n}_{-}$, quantifying the amount by which Inequality 
(\ref{symppt}) is violated. Thus, for our aims, 
the eigenvalue $\tilde{n}_{-}$ completely qualifies and quantifies 
the quantum entanglement of a two--mode Gaussian state $\gr{\sigma}$.\par
We finally mention that, as far as symmetric states are concerned, another 
measure of entanglement, the entanglement of formation $E_{F}$ 
\cite{footnote2},
can be actually computed \cite{giedke03}. Fortunately,
since $E_{F}$ turns out to be, again, 
a decreasing function of $\tilde{n}_{-}$, it provides for
symmetric states a quantification
of entanglement fully equivalent to the one provided by the
logarithmic negativity $E_{\N}$. 
Therefore, from now on, 
we will adopt $E_{\N}(\gr{\sigma})$
as the entanglement measure of Gaussian states, recalling
that this quantity constitutes an upper bound to the
{\em distillable entanglement} of quantum states \cite{vidwer}.
\par
\section{Evolution in general Gaussian environments\label{evo}}
We now consider the local evolution of an arbitrary two--mode Gaussian state 
in noisy channels, 
in the presence of arbitrarily squeezed (``phase--sensitive'') environments. 
In general, the two channels related to the two different 
modes could be different from one another,
each mode evolving independently in its channel. 
We will refer to the channel (bath) in which mode $i$ evolves
as to channel (bath) $i$.
The system
is governed, in interaction picture, by the following master equation \cite{walls}
\begin{eqnarray}
\dot \varrho & = & \sum_{i=1,2}\frac{\Gamma}{2}N_{i} \: L[a_{i}^{\dag}]\varrho
+\frac{\Gamma}{2}(N_{i}+1)\:L[a_{i}]\varrho
\nonumber \\ &-&
\frac{\Gamma}{2}\: \Big( \overline{M_{i}}\:D[a_{i}]\varrho + M_{i}
\:D[a_{i}^{\dag}]\varrho \Big)
\label{rhoev} \: ,
\end{eqnarray}
where the dot stands
for time--derivative and
the Lindblad superoperators are defined by
$L[O]\varrho \equiv  2 O\varrho O^{\dag} -
O^{\dag} O\varrho -\varrho O^{\dag} O$ and
$D[O]\varrho \equiv  2 O\varrho O -O O\varrho -\varrho O O$.
The complex parameter $M_{i}$ is the correlation function of bath $i$;
it is usually referred to as the ``squeezing'' of the bath.
$N_{i}$ is instead a phenomenological parameter related to
the purity of the asymptotic stationary state. 
Positivity of the density matrix imposes
the constraint $|M_{i}|^{2} \le N_{i}(N_{i}+1)$.
At thermal equilibrium, {\it i.e.}~for $M_{i}=0$, 
the parameter $N_{i}$ coincides with the average number of 
thermal photons in bath $i$. \par
A squeezed environment, leading to the master equation (\ref{rhoev}), 
may be modeled as the interaction with a bath of 
oscillators excited in squeezed thermal states \cite{sqbath1}. 
Several effective realizations of squeezed baths have been 
proposed in recent years \cite{tomvit,cir}. In particular, 
in Ref.~\cite{tomvit} the authors show that a
squeezed environment can be obtained, for
a mode of the radiation field, by means of 
feedback schemes relying on QND 
`intracavity' measurements, capable of affecting 
the master equation of the system. More specifically,
an effective squeezed reservoir is shown to be 
the result of a continuous homodyne 
monitoring of a field quadrature, 
with the addition of a feedback driving term, coupling the 
homodyne output current with another field quadrature
of the mode.\par 
Let $\varrho_{i}=S(r_{i},\varphi_{i})\nu_{\bar n_{i}}S(r_{i},\varphi_{i})
^{\dag}$ be the environmental Gaussian state of mode $i$ \cite{marian}. 
Here $\bar n_{i}$ denotes the mean number of photons in the thermal 
state $\nu_{\bar n_{i}}$. Its knowledge allows to determine
the purity of the state via the relation 
$\mu_{i}=1/(2\bar n_{i}+1)$. The operator $S(r,\varphi)=\,{\rm exp}\left(
{\frac12 r\,{\rm e}^{-i2\varphi}a^2-\frac12 r\,{\rm e}^{i2\varphi}a^{\dag 2}}
\right)$ is the one--mode squeezing operator. 
A more convenient parametrization of the channel, endowed with 
a direct phenomenological interpretation, can be achieved by expressing 
$N_{i}$ and $M_{i}$ in terms of the three real variables 
$\mu_{i}$, $r_{i}$ and $\varphi_{i}$ \cite{paris03}
\begin{eqnarray}
\mu_{i}&=&\frac{1}{\sqrt{(2N_{i}+1)^{2}-4|M_{i}|^{2}}} 
\: , \label{purasi} \\
&& \nonumber \\
\cosh(2r_{i})&=&\sqrt{1+4\mu_{i}^{2}|M_{i}|^{2}}
\: , \label{squizasi} \\
&& \nonumber \\
\tan(2\varphi_{i})&=&-\tan\left({\rm Arg}{M_{i}}\right)
\: . \label{phiasi}
\end{eqnarray}
Note that the Gaussian state of the environment 
in bath $i$ coincides with the 
asymptotic state of mode $i$, the global asymptotic state being an uncorrelated 
product of the states $\varrho_{i}$'s, 
irrespective of the initial state.\par
With standard techniques, it can be shown that the master equation (\ref{rhoev}) 
corresponds to a Fokker--Planck equation for the Wigner function of the system 
\cite{walls}. 
In compact notation, one has
\begin{equation}
\dot W (X,t)   = 
\frac{\Gamma}{2} \left[\partial_{X}
 \cdot X^{T} \right. 
+
\left.\partial_{X}\,\boldsymbol{\sigma}_{\infty}\,
\partial_{X}^{T}\right]
W(X,t) \; , \label{fokplan}
\end{equation}

\noindent with $\partial_{X}\equiv(\partial_{x_{1}},\partial_{p_{1}},\partial_{x_{2}},
\partial_{p_{2}})$
and with a diffusion matrix 
\begin{equation}
\gr{\sigma}_{\infty}=
\gr{\sigma}_{1\infty}\oplus\gr{\sigma}_{2\infty}=
\left(\begin{array}{cc}
\gr{\sigma}_{1 \infty}&{\bf 0}\\
{\bf 0}&\gr{\sigma}_{2 \infty}
\end{array}
\right)\; ,
\label{canal}
\end{equation} 
resulting from the tensor 
product of the asymptotic Gaussian states $\gr{\sigma}_{i\infty}$'s,
given by
\begin{equation}
\gr{\sigma}_{i\infty}=\left(\begin{array}{cc}
\frac12 +N_{i}+\,{\rm Re}\,M_{i} & \,{\rm Im}\,M_{i} \\
\,{\rm Im}\,M_{i} & \frac12 +N_{i}-\,{\rm Re}\,M_{i}
\end{array}\right) \, . \label{canalino}
\end{equation}
For an initial Gaussian state of the form Eq.~(\ref{wigner}), 
the Fokker--Planck equation (\ref{fokplan}) 
corresponds to a set of decoupled equations for the second moments and can be easily solved. 
Note that the drift term always damps to $0$ the first statistical moments, and it may
thus be neglected for our aims.
The evolution in the bath preserves the Gaussian form of the initial condition and 
is described by the following equation for the covariance matrix \cite{duan97,paris03,kim95}
\begin{equation}
\gr{\sigma}(t)=\gr{\sigma}_{\infty}
\left(1-{\rm e}^{-\Gamma t}\right)
+\gr{\sigma}(0)\,{\rm e}^{-\Gamma t}.
\label{solution}
\end{equation}
This is a simple Gaussian completely positive map, and 
$\gr{\sigma}(t)$ satisfies the uncertainty relation Eq.~(\ref{bonfide}) if and only if 
the latter is satisfied by both $\gr{\sigma}_{\infty}$ and 
$\gr{\sigma}_{0}$. The compliance 
of $\gr{\sigma}_{\infty}$ with inequality Eq.~(\ref{bonfide}) 
is equivalent to the conditions $|M_{i}|\le N_{i}(N_{i}+1)$.\par
It is easy to see that Eq.~(\ref{solution}) describes 
the evolution of an initial Gaussian state $\gr{\sigma}_{0}$ in an 
arbitrary Gaussian environment $\gr{\sigma}_{\infty}$, which can in general be 
different from that defined by Eq.~(\ref{canal}). It would be interesting to find systems 
whose dynamics could be effectively described by the dissipation in a 
correlated Gaussian environment (recall that the instance we are analyzing involves a 
completely uncorrelated environment). Some perspectives in this 
direction, that lie outside the scopes of the present paper,
could come from feedback and conditional measurement schemes.\par
The initial Gaussian state is described, in general, by a set of ten covariances. To 
simplify the problem and to better point out the relevant features of the non--unitary 
evolution, we will choose an initial state already brought
in standard form: $\gr{\sigma}_{0}=\gr{\sigma}_{sf}$. With this choice
the parametrization of the initial state is completely determined
by the four parameters $a$, $b$, $c_{1}$ and $c_{2}$, defined in Eq.~(\ref{stform}).
This choice is not restrictive as far as the dynamics of purity and
entanglement are concerned.
In fact, let us consider the most 
general initial Gaussian state $\gr{\sigma}$ evolving in the most general 
Gaussian uncorrelated environment $\oplus_{i}\gr{\sigma}_{i \infty}$. 
The state $\gr{\sigma}$ can always be put in standard form
by means of some local transformation $S_{l}=\oplus_{i}S_{i}$. 
Under the same transformation, the state of the environment 
$\oplus_{i}\gr{\sigma'}_{i \infty}$ remains 
uncorrelated, with $\gr{\sigma'}_{i \infty}=S_{i}^{T}\gr{\sigma}_{i \infty}S_{i}$. 
All the properties of entanglement and mixedness for the evolving state are invariant 
under local operations. Therefore, we can state that the evolution of the mixedness and 
of the entanglement of any initial Gaussian state $\gr{\sigma}$ in any uncorrelated 
Gaussian environment $\gr{\sigma}_{\infty}$ is equivalent to the evolution 
of the initial state in standard form $S_{l}^{T}\gr{\sigma}S_{l}$
in the uncorrelated Gaussian environment $S_{l}^{T}\gr{\sigma}_{\infty}S_{l}$.\par
Finally, to further simplify the dynamics of the state
without loss of generality, we can set $\varphi_{1}=0$ ( 
corresponding to $\,{\rm Im}\,M_{1}=0$) as a ``reference choice'' for phase space
rotations.\par
Quite obviously, the standard form of the state is not preserved in an arbitrary channel, as can
be seen from Eqs.~(\ref{canalino}) and (\ref{solution}). \par
\section{Evolution of mixedness and entanglement}
Let us now consider the evolution of mixedness and entanglement of a generic state 
in standard form (parametrized by $a$, $b$, $c_{1}$ and $c_{2}$) in a generic channel
(parametrized by $\mu_{1}$, $r_{1}$, $\mu_{2}$, $r_{2}$ and $\varphi_{2}$). 
Knowledge of the exact evolution of the state in the channel, given by Eq.~(\ref{solution}), 
allows to apply the results reviewed in Sec.~\ref{sec2m} to keep track of the 
quantities $\mu(t)$, $S_{V}(t)$, $I(t)$ and $E_{\N}(t)$ during the 
non-unitary evolution in the channel. However the explicit dependence of such quantities 
on the nine parameters characterizing the initial state and the environment is quite 
involved. We provide the explicit expressions in App.~\ref{appendix}. They give a systematic 
recipe to compute the evolution of mixedness, correlations and entanglement for
any given Gaussian state in standard form (and, therefore, for any Gaussian state).\par
We now investigate the duration and robustness of  
entanglement during the evolution of the field modes
in the channels. Let us consider an initial entangled state $\gr{\sigma}_{e}$ 
evolving in the bath. 
Making use of the separability criterion Eq.~(\ref{ppt}), 
one finds that the state $\gr{\sigma}_{e}$ becomes separable 
at a certain time $t$ if
\begin{equation}
u\,{\rm e}^{-4\Gamma t}+v\,{\rm e}^{-3\Gamma t}+w\,{\rm e}^{-2\Gamma t}+
y\,{\rm e}^{-\Gamma t}+z=0 \, . \label{dis4}
\end{equation}
The coefficients $u$, $v$, $w$, $y$ and $z$ are functions of the nine parameters 
characterizing the initial state and the channel (see App.~\ref{appendix}). 
Eq.~(\ref{dis4}) is an algebraic equation of fourth degree
in the unknown $k={\rm e}^{-\Gamma t}$. 
The solution $k_{ent}$ of such an equation  
closest to one, and satisfying $k_{ent}\le 1$  
can be found for any given initial 
entangled state. Its knowledge promptly leads to the determination 
of the ``entanglement time'' $t_{ent}$ of
the initial state in the channel, defined as the time interval after which the initial state 
becomes separable
\begin{equation}
t_{ent}=-\frac1\Gamma \ln k_{ent} \, .
\label{loga}
\end{equation} \par
\begin{figure}[tb]
\begin{center}
\psfig{file=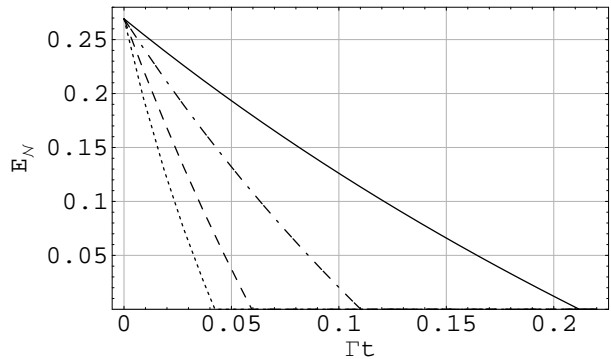,width=8cm}
\caption{Time evolution of logarithmic negativity of a non symmetric Gaussian state with 
$a=2$, $b=1$, $c_{1}=1$, $c_{2}=-1$ in several non correlated environments. 
The solid line refers to the case $\mu_{1}=\mu_{2}=1/2$, $r_{1}=r_{2}=0$;
the dashed line refers to the case $\mu_{1}=1/2$, $\mu_{2}=1/6$, $r_{1}=r_{2}=0$;
the dot--dashed line refers to the case $\mu_{1}=1/6$, $\mu_{2}=1/2$, $r_{1}=r_{2}=0$;
the dotted line refers to the case $\mu_{1}=\mu_{2}=1/2$, $r_{1}=r_{2}=1$. In all
cases the squeezing angle $\varphi_{2}=0$. All the plotted quantities are dimensionless.
\label{entgningn}}
\end{center}
\end{figure}
\begin{figure}[tb]
\psfig{file=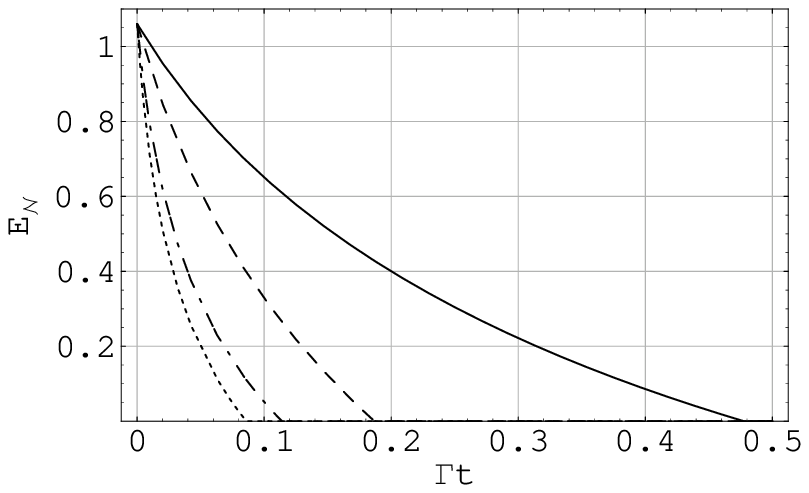,width=8cm}
\caption{Time evolution of logarithmic negativity of a symmetric Gaussian state with 
$a=1.5$, $b=1.5$, $c_{1}=1.2$, $c_{2}=-1.4$ in several non correlated environments. 
The solid line refers to the case $\mu_{1}=\mu_{2}=1/2$, $r_{1}=r_{2}=0$;
the dashed line refers to the case $\mu_{1}=\mu_{2}=1/4$, $r_{1}=r_{2}=0$;
the dot--dashed line refers to the case $\mu_{1}=\mu_{2}=1/2$, $r_{1}=r_{2}=1$;
the dotted line refers to the case $\mu_{1}=\mu_{2}=1/2$, $r_{1}=0$, $r_{2}=1.5$.  
In all cases the squeezing angle $\varphi_{2}=0$. All the plotted quantities are dimensionless.
\label{entsyingn}}
\end{figure}
\begin{figure}[tb]
\psfig{file=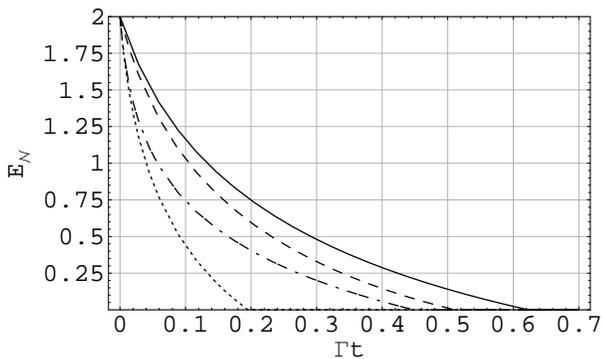,width=8cm}
\caption{Time evolution of logarithmic negativity of a two--mode squeezed state with 
$r=1$ in several non correlated environments. 
The solid line refers to the case $\mu_{1}=\mu_{2}=1/2$, $r_{1}=r_{2}=0$, $\varphi_{2}=0$;
the dashed line refers to the case $\mu_{1}=4$, $\mu_{2}=1$, $r_{1}=r_{2}=0$, $\varphi_{2}=0$;
the dot--dashed line refers to the case $\mu_{1}=\mu_{2}=1/2$, $r_{1}=r_{2}=1$, $\varphi_{2}=0$;
the dotted line refers to the case $\mu_{1}=\mu_{2}=1/2$, $r_{1}=r_{2}=1$, $\varphi_{2}=\pi/4$. 
All the plotted quantities are dimensionless.
\label{ent2m}}
\end{figure}
\begin{figure}[tb]
\psfig{file=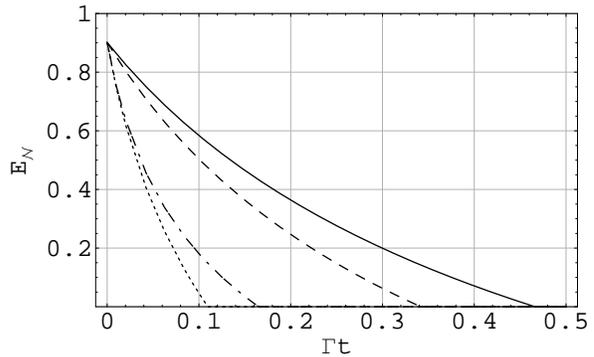,width=8cm}
\caption{Time evolution of logarithmic negativity of a two--mode squeezed thermal state with 
initial purity $\mu=1/9$, $r=1$ in several non correlated environments. 
The solid line refers to the case $\mu_{1}=\mu_{2}=1/2$, $r_{1}=r_{2}=0$, $\varphi_{2}=0$;
the dashed line refers to the case $\mu_{1}=4$, $\mu_{2}=1$, $r_{1}=r_{2}=0$, $\varphi_{2}=0$;
the dot--dashed line refers to the case $\mu_{1}=\mu_{2}=1/2$, $r_{1}=r_{2}=1$, $\varphi_{2}=0$;
the dotted line refers to the case $\mu_{1}=\mu_{2}=1/2$, $r_{1}=r_{2}=1$, 
$\varphi_{2}=\pi/4$. All the plotted quantities are dimensionless.
\label{ent2mth}}
\end{figure}
\begin{figure}[tb]
\psfig{file=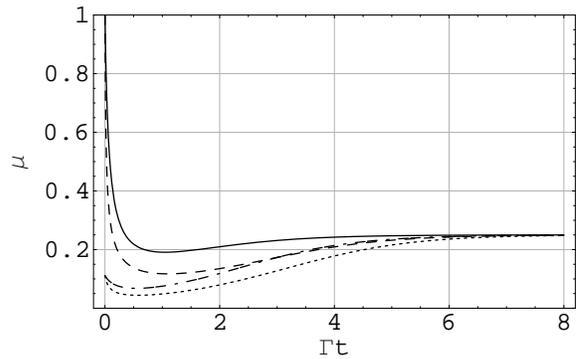,width=8cm}
\caption{Time evolution of the purity of two--mode squeezed thermal states.
The solid line refers to a two--mode squeezed vacuum state with $r=1$
in an environment with $\mu_{1}=\mu_{2}=1/2$, $r_{1}=r_{2}=0$; the 
dashed line shows the behavior of the same state for 
$\mu_{1}=\mu_{2}=1/2$, $r_{1}=r_{2}=1$. The dot--dashed line refers to a 
mixed state with $\mu=1/9$, $r=1$ in an environment with 
$\mu_{1}=\mu_{2}=1/2$, $r_{1}=r_{2}=0$; the dotted line refers to the same 
state for $\mu_{1}=\mu_{2}=1/2$, $r_{1}=r_{2}=1$. The squeezing angle 
$\varphi_{2}$ has always been set to $0$. 
All the plotted quantities are dimensionless.
\label{pur2m}}
\end{figure}
\begin{figure}[tb]
\psfig{file=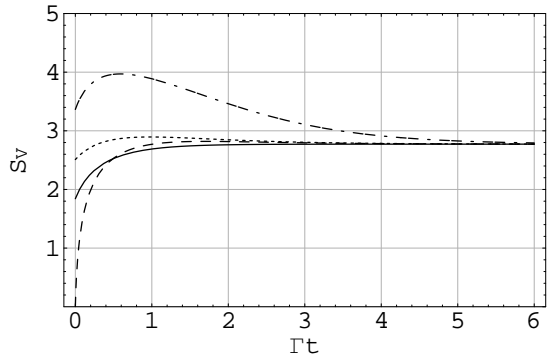,width=8cm}
\caption{Time evolution of the Von Neumann entropy of several 
Gaussian states in a thermal environment with $\mu_{1}=\mu_{2}=1/3$. 
the solid line refers to a state with $a=1$, $b=c_{1}=-c_{2}=1$; the 
dashed line refers to a two--mode squeezed vacuum state with $r=1$; the 
dot--dashed line refers to a squeezed thermal state with $\mu=1/16$, $r=1$; 
the dotted line refers to a non entangled state with $a=b=2$, 
$c_{1}=-c_{2}=1.5$. The squeezing angle 
$\varphi_{2}$ has always been set to $0$. 
All the plotted quantities are dimensionless.
\label{vneufig}}
\end{figure}
\begin{figure}[tb]
\psfig{file=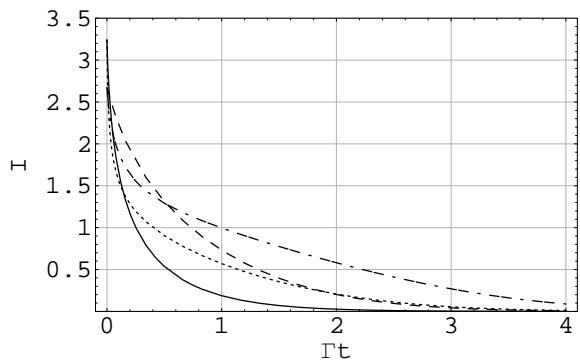,width=8cm}
\caption{Time evolution of the mutual information of two--mode squeezed thermal states 
in an environment with $\mu_{1}=\mu_{2}=1/3$. The solid line refers to a pure state 
with $r=1$ in a non squeezed environment; the dotted line refers to the same state 
in an environment with $r_{1}=r_{2}=1$; the dashed line refers to a squeezed thermal state 
with $\mu=1/16$, $r=1$ in a non squeezed environment; the dot--dashed line refers 
to the same state in a squeezed environment with $r_{1}=r_{2}=1$. The squeezing angle 
$\varphi_{2}$ has always been set to $0$. 
All the plotted quantities are dimensionless.
\label{minf2m}}
\end{figure}
\begin{figure}[t]
\psfig{file=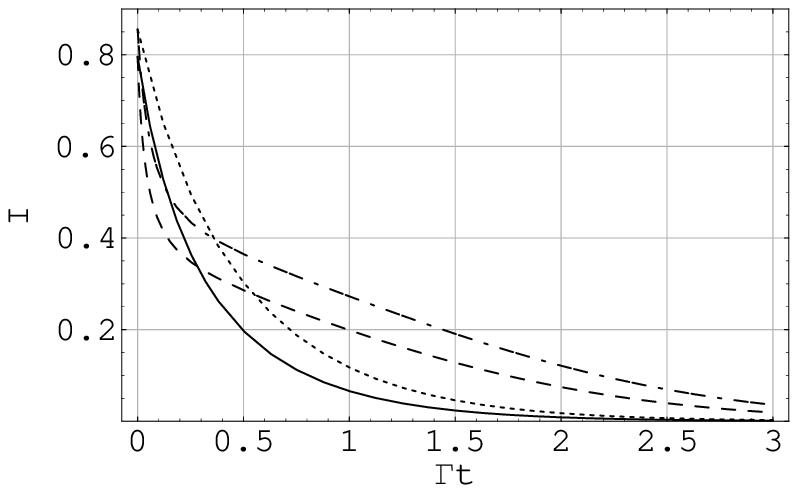,width=8cm}
\caption{Time evolution of the mutual information of Gaussian states 
in an environment with $\mu_{1}=\mu_{2}=1/3$. The solid line refers to state
with $a=2$, $b=c_{1}=-c_{2}=1$ in a non squeezed environment; 
the dotted line refers to the same state 
in an environment with $r_{1}=r_{2}=1$; the dashed line refers to a 
non entangled state 
with $a=b=2$, $c_{1}=-c_{2}=1.5$ in a non squeezed environment; the dot--dashed line refers 
to the same state in a squeezed environment with $r_{1}=r_{2}=1$. The squeezing angle 
$\varphi_{2}$ has always been set to $0$. 
All the plotted quantities are dimensionless.
\label{minfgn}}
\end{figure}
The results of the numerical analysis of the evolution of entanglement and mixedness
for several initial states are reported 
in Figs.~\ref{entgningn} through \ref{minfgn}.
In general, one can see that, trivially, a less mixed environment better preserves 
both purity and entanglement by prolonging the entanglement time. More remarkably,   
Fig.~\ref{entgningn} shows that local squeezing of the two 
uncorrelated channels
does not help preserving quantum correlations between the evolving modes. Moreover,
as can be seen from Fig.~\ref{entgningn}, 
states with greater uncertainties on, say, mode $1$ ($a>b$) 
better preserves its entanglement if bath $1$ is more mixed 
than bath $2$ ($\mu_{1}<\mu_{2}$). A quite interesting feature 
is shown in Fig.~\ref{minfgn}: the mutual information 
is better preserved in squeezed channels, especially at long times. 
This property has been tested 
as well on non entangled states, endowed only with classical 
correlations, see Fig.~\ref{minfgn}, and on 
two--mode squeezed states, see Fig.~\ref{minf2m}, and seems to hold 
quite generally.
In Fig.~\ref{entsyingn}, the behavior of some initially symmetric
states is considered. In this instance we can see that, in squeezed baths, 
the entanglement of the initial state is better preserved if the 
squeezing of the two channels is balanced.\par
The analytic optimization of the relevant quantities characterizing 
mixedness and correlations in the channel turn out to be difficult in the general 
case. Thus it is convenient to proceed with our analysis by dealing with
particular instances of major phenomenological interest.
\subsection{Standard form states in thermal channels}
In this subsection, we deal with the case of states in generic standard form 
(parametrized by $a$, $b$, $c_{i}$) evolving in
two thermal channels (parametrized by two --
generally different -- mean photon numbers $N_{i}$'s). This instance is particularly 
relevant, because it gives a basic description of actual experimental settings 
involving, for instance, fiber--mediated communications. \par
The purity $\mu$ of the global quantum state turns out to be a decreasing 
function of the $N_{i}$'s at any given time. 
The symplectic eigenvalue $\tilde{n}_{-}$ is also in general 
an increasing function of the $N_{i}$'s. Therefore,
the entanglement of the evolving state is optimal for ideal 
vacuum environments, which is quite trivial, recalling the well understood 
synergy between entanglement and purity for general quantum states.
\par
\subsection{Entanglement time of symmetric states}
We have already provided a way of computing the entanglement 
time of an arbitrary two--mode state in arbitrary channels. The expression 
of such a quantity is, unfortunately, rather involved in the general case. 
However, focusing on symmetric states (which satisfy $a=b$), some 
simple analytic results can be found, thanks to the simple form 
taken by $\tilde{n}_{-}$ for these states. 
An initially symmetric entangled state maintains its symmetric standard form 
if evolving in equal, 
independent environments (with $N_{1}=N_{2}\equiv N_{B}$). 
This is the instance we will consider in the following. \par
Let us suppose that $|c_{1}|\le|c_{2}|$, then 
Eqs.~(\ref{symppt}) and (\ref{symeig}) provide
the following bounds for the entanglement time
\be
\ln\left(1+
\frac{|c_{1}|-a+\frac12}{N_{B}}\right)\le\Gamma t_{ent}\le
\ln\left(1+
\frac{|c_{2}|-a+\frac12}{N_{B}}\right) \, . \label{tentbnd}
\ee
Imposing the additional property $c_{1}=-c_{2}$ 
we obtain standard forms which can be written as squeezed thermal states
(see Eqs.~\ref{sqthe}). 
For such states, Inequality Eq.~(\ref{tentbnd}) reduces to
\be
t_{ent}=\frac{1}{\Gamma}\ln\left(1+
\frac{1-\,{\rm e}^{-2r}}{2\sqrt{\mu}N_{B}}\right)\; .
\ee
In particular, for $\mu=1$, one recovers the entanglement 
time of a two--mode squeezed vacuum state in a thermal channel 
\cite{duan00,paris02, prauz03}. Note that two--mode 
squeezed vacuum states encompass all the possible standard forms of 
pure Gaussian states.
\subsection{Two--mode squeezed thermal states}
As we have seen,
two--mode squeezed thermal states constitute 
a relevant class of Gaussian states, parametrized 
by their purity $\mu$ and by the squeezing parameter $r$
according to Eqs.~(\ref{sqthe}).
In particular, two--mode squeezed vacuum states
(or twin-beams), which can be defined as 
squeezed thermal states with $\mu=1$, 
correspond to maximally 
entangled symmetric states for fixed marginal purity. 
Therefore, they constitute a crucial resource
for possible applications of Gaussian states in quantum 
information engineering.\par
For squeezed thermal states (chosen as initial conditions 
in the channel), it can be shown analytically that the 
partially transposed symplectic eigenvalue $\tilde{n}_{-}$
is at any time an increasing function of the bath squeezing angle
$\varphi_{2}$: ``parallel'' squeezing in the two channels
optimizes the preservation of entanglement. 
Both in the instance of two equal squeezed baths ({\it i.e.~}with 
$r_{1}=r_{2}=r$) and of a thermal bath joined to
a squeezed one ({\it i.e.~}$r_{1}=r$ and $r_{2}=0$),
it can be shown that $\tilde{n}_{-}$ is an  
increasing function of $r$. 
These analytical results agree with those provided
in Ref.~\cite{kim03} in the study of the qualitative
degradation of entanglement for pure squeezed states.
The proofs of the above statements are sketched in App.~\ref{appb}.\par 
Such analytical considerations, supported by direct 
numerical analysis, clearly show that
a local squeezing of the environment faster
degrades the entanglement of the initial state. The same behavior 
occurs for purity. The time evolution of the logarithmic negativity
of two--mode squeezed states -- thermal and pure --  is shown 
in Figs.~\ref{ent2m} and \ref{ent2mth}.
The evolution of the global purity 
is reported in Fig.~\ref{pur2m}, while the evolutions of the
Von Neumann entropy and of the mutual information are shown, respectively,
in Figs.~\ref{vneufig} and \ref{minf2m}.
\section{Summary and conclusions}
We studied the evolution of mixedness, entanglement and mutual information
of initial two--mode Gaussian states evolving in uncorrelated Gaussian 
environments. We derived exact general relations that allow to determine
the time evolution of such quantities, and provided
analytical estimates on the entanglement time. 
The optimal bath parameters for the preservation of quantum correlations 
and purity have been determined for thermal baths and for two--mode squeezed states 
in more general baths. A detailed numerical analysis has been performed 
for the most general cases. \par
We found that, in general, a local squeezing of the baths does not help 
to preserve purity and quantum correlations of the evolving state, 
both at small times ({\em i.e.~}for $\Gamma t\lesssim 1$) and asymptotically. 
On the other hand, local squeezing of the baths can improve 
the preservation of the mutual information in uncorrelated channels.
Besides, coherence and correlations are better maintained in 
environments with lower average number of photons.\par
The present study may be be extended to the case of $n$-mode 
Gaussian states. This generalization would be desirable, 
since the practical implementation of quantum information protocols 
usually requires some redundancy. For three--mode Gaussian states, 
separability conditions analogous to Inequality Eq.~(\ref{sepcomp}) have 
been determined \cite{giedke01}, and could be exploited to provide a
qualitative picture of the evolution of three--mode entanglement in
noisy channels.  
\subsection*{Acknowledgments}
AS, FI and SDS
thank INFM, INFN, and MIUR under national project PRIN-COFIN 2002
for financial support. 
The work of MGAP is supported in part by UE programs ATESIT 
(Contract No. IST-2000-29681). 
MGAP is a research fellow at {\em Collegio Volta}.
\appendix
\section{Explicit determination of mixedness and entanglement in the general
case\label{appendix}}
Here we provide explicit expressions which allow to determine the exact evolution
in uncorrelated channels of a generic initial state in standard form. 
The relevant quantities $E_{\N}$, $\mu$, $S_{V}$, $I$, as we have seen,
are all functions of the four $Sp_{(2,\mathbb{R})}\oplus Sp_{(2,\mathbb{R})}$
invariants. Let us then write such quantities as follows
\bea
{\rm Det}\,\gr{\sigma}&=&\sum_{k=0}^{4}\Sigma_{k}\,{\rm e}^{-k\Gamma t}\; ,\\
{\rm Det}\,\gr{\alpha}&=&\sum_{k=0}^{2}\alpha_{k}\,{\rm e}^{-k\Gamma t}\; ,\\
{\rm Det}\,\gr{\beta}&=&\sum_{k=0}^{2}\beta_{k}\,{\rm e}^{-k\Gamma t}\; ,\\
{\rm Det}\,\gr{\gamma}&=&\gamma_{2}\,{\rm e}^{-2\Gamma t}\; ,\\
\eea
defining the sets of coefficients $\Sigma_{i}$, $\alpha_{i}$, $\beta_{i}$, 
$\gamma_{i}$. One has
\begin{widetext}
\bea
\Sigma_{4}&=&a^2 b^2+\frac{a^2}{4\mu^2_{2}}+\frac{b^2}{4\mu^2_{1}}
-a^2 b\frac{\cosh2r_{2}}{\mu_{2}}-a b^2\frac{\cosh2r_{1}}{\mu_{1}}
+ab\frac{\cosh2r_{1}\cosh2r_{2}}{\mu_{1}\mu_{2}}
-a\frac{\cosh2r_{1}}{4\mu_{1}\mu_{2}^2}-b\frac{\cosh2r_{2}}{4\mu_{1}^2\mu_{2}}
\nonumber \\
&&+(c_{1}^2+c_{2}^2)\left(a\frac{\cosh2r_{2}}{2\mu_{2}}
+\frac{b\cosh2r_{1}}{2\mu_{1}}
-\frac{\cosh2r_{1}\cosh2r_{2}}{4\mu_{1}\mu_{2}}-
\frac{\sinh2r_{1}\sinh2r_{2}\cos2\varphi_{2}}{4\mu_{1}\mu_{2}}-ab\right)
\nonumber \\
&&+(c_{1}^2-c_{2}^2)\left(a\frac{\sinh2r_{2}\cos2\varphi_{2}}{2\mu_{2}}
+b\frac{\sinh2r_{1}}{2\mu_{1}}
-\frac{\sinh2r_{1}\cosh2r_{2}}{4\mu_{1}\mu_{2}}
-\frac{\cosh2r_{1}\sinh2r_{2}\cos2\varphi_{2}}{4\mu_{1}\mu_{2}}\right)
\nonumber \\
&&+c_{1}^2c_{2}^2+\frac{1}{16\mu_{1}^2 \mu_{2}^2} \; , \\
&&\nonumber\\
\Sigma_{3}&=&-2\frac{a^2}{4\mu^2_{2}}-2\frac{b^2}{4\mu^2_{1}}
+a^2 b\frac{\cosh2r_{2}}{\mu_{2}}+a b^2\frac{\cosh2r_{1}}{\mu_{1}}
-2ab\frac{\cosh2r_{1}\cosh2r_{2}}{\mu_{1}\mu_{2}}
+3a\frac{\cosh2r_{1}}{4\mu_{1}\mu_{2}^2}+3b\frac{\cosh2r_{2}}{4\mu_{1}^2\mu_{2}}
\nonumber\\
&&-(c_{1}^2-c_{2}^2)\left(a\frac{\sinh2r_{2}\cos2\varphi_{2}}{2\mu_{2}}
+b\frac{\sinh2r_{1}}{2\mu_{1}}
-2\frac{\sinh2r_{1}\cosh2r_{2}}{4\mu_{1}\mu_{2}}
-2\frac{\cosh2r_{1}\sinh2r_{2}\cos2\varphi_{2}}{4\mu_{1}\mu_{2}}\right)
\nonumber\\
&&-(c_{1}^2+c_{2}^2)\left(a\frac{\cosh2r_{2}}{2\mu_{2}}
+\frac{b\cosh2r_{1}}{2\mu_{1}}
-2\frac{\cosh2r_{1}\cosh2r_{2}}{4\mu_{1}\mu_{2}}-2
\frac{\sinh2r_{1}\sinh2r_{2}\cos2\varphi_{2}}{4\mu_{1}\mu_{2}}\right)
-\frac{1}{4\mu_{1}^2 \mu_{2}^2} \, , \\
&&\nonumber\\
\Sigma_{2}&=&\frac{a^2}{4\mu^2_{2}}+\frac{b^2}{4\mu^2_{1}}
+ab\frac{\cosh2r_{1}\cosh2r_{2}}{\mu_{1}\mu_{2}}
-3a\frac{\cosh2r_{1}}{4\mu_{1}\mu_{2}^2}
-3b\frac{\cosh2r_{2}}{4\mu_{1}^2\mu_{2}}
\nonumber\\
&&-(c_{1}^2+c_{2}^2)\left(
\frac{\cosh2r_{1}\cosh2r_{2}}{4\mu_{1}\mu_{2}}+
\frac{\sinh2r_{1}\sinh2r_{2}\cos2\varphi_{2}}{4\mu_{1}\mu_{2}}\right)
\nonumber \\
&&-(c_{1}^2-c_{2}^2)\left(
\frac{\sinh2r_{1}\cosh2r_{2}}{4\mu_{1}\mu_{2}}
+\frac{\cosh2r_{1}\sinh2r_{2}\cos2\varphi_{2}}{4\mu_{1}\mu_{2}}\right)
+\frac{1}{16\mu_{1}^2 \mu_{2}^2} \; , \\
&&\nonumber\\
\Sigma_{1}&=&
+a\frac{\cosh2r_{1}}{4\mu_{1}\mu_{2}^2}+b\frac{\cosh2r_{2}}{4\mu_{1}^2\mu_{2}}
-\frac{1}{4\mu_{1}^2 \mu_{2}^2} \; , \\
&&\nonumber\\
\Sigma_{0}&=&\frac{1}{16\mu_{1}^2\mu_{2}^2} \; ,
\eea
\end{widetext}
\bea
\alpha_{2}&=&a^2-a\frac{\cosh2r_{1}}{\mu_{1}}+\frac{1}{4\mu_{1}^2}\; ,\\
\alpha_{1}&=&a\frac{\cosh2r_{1}}{\mu_{1}}-2\frac{1}{4\mu_{1}^2}\; ,\\
\alpha_{0}&=&\frac{1}{4\mu_{1}^2}\; , \\
\beta_{2}&=&b^2-b\frac{\cosh2r_{2}}{\mu_{2}}+\frac{1}{4\mu_{2}^2}\; ,\\
\beta_{1}&=&b\frac{\cosh2r_{2}}{\mu_{2}}-2\frac{1}{4\mu_{2}^2}\; ,\\
\beta_{0}&=&\frac{1}{4\mu_{2}^2}\; , \\
\gamma_{2}&=&c_{1}c_{2}\; .
\eea

The coefficients of Eq.~(\ref{dis4}), 
whose solution $k_{ent}$ allows to determine the entanglement time 
of an arbitrary two--mode Gaussian state, read
\bea
u&=&\Sigma_{4}\; ,\\
v&=&\Sigma_{3}\; ,\\
w&=&\Sigma_{2}-\alpha_{2}-\beta_{2}-|\gamma_{2}|\; ,\\
y&=&\Sigma_{1}-\alpha_{1}-\beta_{1}\; ,\\
z&=&\Sigma_{0}-\alpha_{0}-\beta_{0}+\frac14\; .
\eea 

\section{Proofs for two--mode squeezed states\label{appb}}

In this appendix we consider a two--mode squeezed thermal state of the 
form of Eq.~(\ref{sqthe}) as the initial input in the noisy channels. \par
We first deal with the dependence of entanglement and 
mixedness on the squeezing angle $\varphi_{2}$ of bath $2$. 
It can be easily shown (see App.~\ref{appendix}) that 
$\Delta(\gr{\sigma})$ does not depend on $\varphi_{2}$, whereas
${\rm Det}\,\gr{\sigma}$ turns out to be a decreasing function of 
$\cos\,\varphi_{2}$. Therefore, since the symplectic eigenvalue
$\tilde{n}_{-}$ increases with ${\rm Det}\,\gr{\sigma}$, one has  
that $\varphi_{2}=0$ is the optimal choice for maximizing 
both entanglement and purity of the evolving state.\par
We now address the instance of two equally squeezed baths, with 
$N_{i}\equiv N_{B}$, $r_{i}\equiv r_{B}$ and 
$\varphi_{2}=0$. The time dependent covariance matrix
$\gr{\sigma}_{2m}$ can be 
written in the form
\[
\boldsymbol{\sigma}_{2m}
= \left(\begin{array}{cccc}
j_{-}&0&k&0\\
0&j_{+}&0&-k\\
k&0&j_{-}&0\\
0&-k&0&j_{+}
\end{array}\right)\; , 
\]
with
\bea
j_{\mp}&=&\frac{\cosh2r}{2\sqrt{\mu}}\,{\rm e}^{-\Gamma t} 
+ (N_{B}+\frac12)\,{\rm e}^{\mp2r_{B}}(1-\,{\rm e}^{-\Gamma t})  ,\nonumber \\
k&=&\frac{\sinh2r}{2\sqrt{\mu}}\,{\rm e}^{-\Gamma t} \; . \nonumber
\eea
The standard form of $\gr{\sigma}_{2m}$ is easily found just by squeezing 
the field in the two modes of the same quantity $\sqrt{j_{+}/j_{-}}$. The result 
is a symmetric standard form, whose smallest partially transposed symplectic 
eigenvalue $\tilde{n}_{-}$ can be computed according to Eq.~(\ref{symeig})
\[
\tilde{n}_{-}=(j_{-}-k)(j_{+}-k) =
d\cosh2r_{B}+\ldots\; ,
\]
where the terms that do not depend on $r_{B}$ are irrelevant to our discussion
and have thus been neglected.
The coefficient $d$ is a positive function of $t$, $r$ and $N_{B}$, so that 
the best choice to maximize entanglement at any given time is given by $r_{B}=0$.
Quite obviously, $\tilde{n}_{-}$ turns out to be an increasing function of $N_{B}$ 
as well.\par
Finally, we deal with the instance in which bath 1 is squeezed while bath 2 is thermal, 
with $r_{2}=0$. For ease of notation we define $|\gr{\sigma}|=\,{\rm Det}\,
\gr{\sigma}$. We recall that $2\tilde{n}^{2}_{-}=\tilde{\Delta}- 
\sqrt{\tilde{\Delta}^2 -4|\gr{\sigma}|}$. Thus, for entangled states
(for which $\tilde{n}_{-}<1/2$), one finds
\[
\partial_{|\gr{\sigma}|}(2\tilde{n}^{2}_{-})>
-4\partial_{\tilde{\Delta}}(2\tilde{n}^2_{-})>0\; .
\] 
The sign of the quantity $4\partial_{r_{1}}|\gr{\sigma}|-
\partial_{r_{1}}\tilde{\Delta}$ for the case of the initial 
two--mode squeezed can be shown, after some algebra, to 
be determined by 
\[
4(\,{\rm e}^{\Gamma t}-1)\cosh2r n_{2}^2 + (3+\cosh4r)n_{2} -
(\,{\rm e}^{\Gamma t}+1)\cosh2r \, . 
\] 
This second degree polynomial is positive for 
$n_{2}\equiv N_2+1/2\ge 1/2$. This proves that the entanglement decreases 
as the squeezing of bath 1 increases.



\begin{thebibliography}{99}

\bibitem{pati03} {\em Quantum Information Theory with Continuous Variables},
S. L. Braunstein and A. K. Pati Eds. (Kluwer, Dordrecht, 2002).

\bibitem{crypto} H. P. Yuen and A. Kim, Phys. Lett. A {\bf 241}, 
135 (1998);
F. Grosshans, G. Van Assche, J. Wenger, R. Brouri, N. J. Cerf,
and P. Grangier, Nature {\bf 421}, 238 (2003).

\bibitem{tele} A. Furusawa, J. L. Sorensen, S. L. Braunstein, C. A.
Fuchs, H. J. Kimble, and E. S. Polzik, Science {\bf 282}, 706 (1998); 
T. C. Zhang, K. W. Goh, C. W. Chou, P. Lodahl, and H. J. Kimble,
Phys. Rev. A {\bf 67}, 033802 (2003).

\bibitem{duan97} L.-M. Duan and G.-C. Guo, 
Quantum Semiclass. Opt. {\bf 9}, 953 (1997).

\bibitem{hiroshima01} T. Hiroshima, Phys. Rev. A {\bf 63}, 022305 (2001).

\bibitem{scheel01} S. Scheel and D.-G. Welsch, 
Phys. Rev. A {\bf 64}, 063811 (2001).

\bibitem{paris02} M. G. A Paris, {\em Entangled Light and Applications} 
in {\em Progress in Quantum Physics Research}, edited by V. Krasnoholovets ,  
(Nova Publisher, New York, in press).

\bibitem{kim03} D. Wilson, J. Lee, and M. S. Kim, J. Mod. Opt. {\bf 50}, 1809
(2003).

\bibitem{paris04} S. Olivares, M. G. A. Paris, and A. R. Rossi, Phys.
Lett. A {\bf 319}, 32 (2003).

\bibitem{prauz03} J. S. Prauzner--Bechcicki, e--print quant--ph/0211114 (2003).

\bibitem{bowen03} W. P. Bowen, R. Schnabel, P. K. Lam, and T. C. Ralph, 
Phys. Rev. Lett. {\bf 90}, 043601 (2003).

\bibitem{vidwer} G. Vidal and R. F. Werner, Phys. Rev. A {\bf 65}, 032314 (2002).

\bibitem{tomvit} P. Tombesi and D. Vitali, 
Phys. Rev. A {\bf 50}, 4253 (1994); P. Tombesi, and D. Vitali, 
Appl. Phys. B {\bf 60}, S69 (1995).

\bibitem{cir}N. L\"utkenhaus, J. I. Cirac, and P. Zoller, Phys. Rev. A {\bf 57}, 
548 (1998).

\bibitem{wise}H. M. Wiseman and G. J. Milburn, Phys. Rev. Lett. {\bf 70}, 548 (1993); 
H. M. Wiseman and G. J. Milburn, Phys. Rev. A {\bf 49}, 1350 (1994).

\bibitem{simon00} R. Simon, Phys. Rev. Lett. {\bf 84}, 2726 (2000).

\bibitem{barnett} See, e.g., S. M. Barnett and P. M. Radmore, 
{\em Methods in Theoretical Quantum Optics} 
(Clarendon Press, Oxford, 1997).

\bibitem{simon8794} R. Simon, N. Mukunda, and B. Dutta, Phys. Rev. A {\bf 49}, 1567 (1994).

\bibitem{duan00} L.-M. Duan, G. Giedke, J. I. Cirac, and 
P. Zoller, Phys. Rev. Lett. {\bf 84}, 2722 (2000).

\bibitem{holevo} A. S. Holevo, M. Sohma, and O. Hirota, Phys. Rev. A {\bf 59}, 1820 (1999); 
A. S. Holevo and R. F. Werner, ibid. {\bf 63}, 032312 (2001).

\bibitem{serafozzi} A. Serafini, F. Illuminati, and S. De Siena, J. Phys. B: At. Mol. Op.
Phys. {\bf 37}, L21 (2004).

\bibitem{2max} S. M. Barnett, S. J. D. Phoenix,  Phys. Rev. A {\bf 40}, 2404
(1989); S. M. Barnett, S. J. D. Phoenix,  ibid. {\bf 44}, 535 
(1991); M. G. A. Paris, ibid. {\bf 59}, 1615 (1999).

\bibitem{giedke03} G. Giedke, M. M. Wolf, O. Kr\"uger, 
R. F. Werner, and J. I. Cirac, Phys. Rev. Lett. {\bf 91}, 107901 (2003).

\bibitem{agarwal71} G. S. Agarwal, Phys. Rev. A {\bf 3}, 828 (1971).

\bibitem{vedral01} L. Henderson and V. Vedral, 
J. Phys. A {\bf 34}, 6899 (2001).

\bibitem{footnote1} The proof of such properties can be found in Ref.~\cite{vidwer}
for finite dimensional Hilbert spaces. Anyway, since the proof essentially relies on 
the use of the Jordan decomposition, it still holds in infinite 
dimension as shown in Ref.~\cite{eiserth}.

\bibitem{eiserth} J. Eisert, PhD thesis, University of Potsdam (Potsdam, 2001).

\bibitem{jensplenio} J. Eisert, C. Simon, and M. B. Plenio, J. Phys. A: 
Math. Gen. {\bf 35}, 3911 (2002).

\bibitem{footnote2} $E_{F}(\varrho)\equiv\min_{\{p_{i},|\psi_{i}\rangle\}}
\sum p_{i}E(|\psi_{i}\rangle\langle\psi_{i}|)$, 
where $E(|\psi\rangle\langle\psi|)$ is the entropy of entanglement 
of the pure state $|\psi\rangle$,
defined as the Von Neumann entropy of its reduced density matrix, 
and the $\min$ is taken over
all the pure states realization of 
$\varrho = \sum p_{i}|\psi_{i}\rangle\langle\psi_{i}|$.

\bibitem{walls} See, e.g., D. Walls and G. Milburn, {\em Quantum Optics}
(Springer Verlag, Berlin, 1994).

\bibitem{sqbath1} Possible squeezed reservoirs are treated in 
M.-A. Dupertuis and S. Stenholm, J. Opt. Soc. Am. B {\bf 4}, 
1094 (1987); M.-A. Dupertuis, S. M. Barnett, and S. Stenholm, ibid. {\bf 4}, 
1102 (1987); Z. Ficek and P. D. Drummond, Phys. Rev. A {\bf 43}, 6247 (1991); 
K. S. Grewal, Phys Rev A {\bf 67}, 022107 (2003);
see also C. W. Gardiner and P. Zoller, {\em Quantum Noise} 
(Springer Verlag, Berlin, 1999) and Ref.~\cite{kim95}.

\bibitem{marian} Any centered single--mode Gaussian state can be written in this
way, see P. Marian and T. A. Marian, Phys. Rev. A {\bf 47}, 4474 (1993).

\bibitem{paris03} M. G. A. Paris, F. Illuminati, A. Serafini, and S. De Siena, Phys. Rev. A 
{\bf 68}, 012314 (2003).

\bibitem{kim95} M. S. Kim and N. Imoto, Phys. Rev. A {\bf 52}, 2401 (1995).

\bibitem{giedke01} G. Giedke, B. Kraus, M. Lewenstein, and J. I. Cirac, Phys. Rev. A {\bf 64}, 
052303 (2001).

\end{thebibliography}
\end{document}